\documentclass{aa}  

\usepackage[utf8]{inputenc}
\usepackage{xcolor}
\usepackage{graphicx}
\usepackage{txfonts}
\usepackage{bbm}
\usepackage{lipsum}
\usepackage{subcaption}         % necessary for continued figures, example in section 3
\usepackage{lscape}             % to rotate a single page table, example in appendix.
\usepackage{placeins}           % useful with \FloatBarrier, to keep 
\graphicspath{{bilder/}}                                
\usepackage[colorlinks=true,
            linkcolor=red,
            citecolor=violet,
            filecolor=cyan,
            urlcolor=blue]{hyperref}

\usepackage{soul}

\begin{document}

\title{Modeling complex plasma instabilities in space plasmas}

\subtitle{Three-component electron formalism of heat-flux instabilities}

%%%%%%%%%%%%%%%%%%%%%%%%%%%%%%%%%%%%%%%%

%%%%%%%%%%%%%%%%%%%%%%%%%%%%%%%%%%%%%%%%

   \author{Dustin L. Schröder\inst{1}
        \and Marian Lazar\inst{1,2} \and Horst Fichtner\inst{1} \and Rodrigo A. López\inst{3,4} \and Stefaan Poedts\inst{2,5}
        }

   \institute{Institute for Theoretical Physics IV, Ruhr-Universität Bochum, Universitätsstrasse 150, 44780 Bochum, Germany\\
             \email{dustin.schroeder@rub.de }
            \and Centre for mathematical Plasma-Astrophysics, KU Leuven, 3001 Leuven, Belgium
            \and Research Center in the intersection of Plasma Physics, Matter, and Complexity ($P^2 mc$), Comisi\'on Chilena de Energ\'{\i}a Nuclear, Casilla 188-D, Santiago, Chile
            \and Departamento de Ciencias F\'{\i}sicas, Facultad de Ciencias Exactas, Universidad Andres Bello, Sazi\'e 2212, Santiago 8370136, Chile \and Institute of Physics, University of Maria Curie-Sk{\l}odowska, ul.\ Radziszewskiego 10, 20-031 Lublin, Poland\\}

   \date{Received \today}

  \abstract
  % 
  % {}  
   {Despite the fact that electrons observed in situ in space plasmas have three major components -- the quasi-thermal core, suprathermal halo, and strahl-- the analysis of instabilities triggered by kinetic, velocity-space anisotropies (such as relative drifts and temperature anisotropy) generally considers only two of these components.}
   {We aim to demonstrate that a realistic modeling with all three components is achievable in the present analysis focusing on heat-flux instabilities. 
   In the absence of particle--particle collisions, these instabilities are responsible for the regulation of the heat flux carried mainly by suprathermal electrons.} 
   {The velocity distributions of the electron populations were modeled according to in situ observations, with a Maxwellian core and Kappa-distributed halo and strahl components.
   We exploited new advanced numerical codes capable of solving the linear dispersion and stability properties of any plasma system with Maxwellian- and Kappa-distributed populations.}
   {The unstable solutions differ significantly from those obtained with simplified models with only two components (such as core-strahl or core-beam models).
   The growth rates predict the systematic excitation and interplay of two unstable modes, whistler heat-flux and/or fire-hose heat-flux instabilities. {The numerical solver "ALPS" was successfully applied to systems with regularized Kappa distributions, for which the analytical derivation of dispersion relations is not straightforward.}}
   {The two instabilities are triggered by the two relative drifts, core-strahl and halo-strahl, and may have new consequences concerning the regulation of the heat flux.
   Particularly important are the cases when the core-strahl instability known from previous studies is in competition with the new instability driven by the halo-strahl drift, as well as when the two instabilities have the same nature and can accumulate.
   Future studies are thus motivated to confirm these predictions in quasilinear theories and numerical simulations.}


   \maketitle
   \nolinenumbers

%%%%%%%%%%%%%%%%%%%%%%%%%%%%%%%%%%%%%%%%%%%%%%%%%%%%%%%%%%%%%%

%%%%%%%%%%%%%%%%%%%%%%%%%%%%%%%%%%%%%%%%%%%%%%%%%%%%%%%%%%%%%%
%%%%%%%%%%%%%%%%%%%%%%
\section{Introduction}
%%%%%%%%%%%%%%%%%%%%%%

Although seemingly insignificant because of their low mass, electrons can still have major implications in space and astrophysical plasmas. 
Of particular interest are the solar wind (SW) and planetary environments, where in situ observations characterizing electron velocity (or energy) distributions (VDs) are possible.
Three main populations (components) are thus revealed: at low energies, the highly dense, (quasi)thermal core; and at higher energies, the so-called suprathermal halo and strahls/beams aligned to the magnetic field that are intensified in the high-speed SW \citep{Pierrard-etal-2001, Wilson-etal-2019a, Wilson-etal-2019b, Abraham-etal-2022}.
Although they have been highlighted since the first in situ observations \citep{Parker-1958, Olbert1968, Vasyliunas1968, Pilipp-etal-1987}, these suprathermal populations have only been modestly represented in SW modeling for a long time.
More recently, advanced numerical solvers and simulation codes have confirmed the potential for the broad involvement of suprathermal electrons in, for example, the exospheric models of the SW and planetary environments \citep{Maksimovic-etal-1997, Pierrard-2009, Peters-etal-2025}, the main heat flux transport \citep{Pagel-etal-2005, Lazar-etal-2020, Salem-etal-2023}, and plasma radio emissions from the interaction of intense electron beams with the background plasma \citep{Henri-etal-2019, Lazar-etal-2023a}.
In most applications, an important role is played by the electron strahl/beam, whose focus at small pitch angles is reduced with heliocentric distance \citep{Hammond-etal-1996, Bercic-etal-2019}.
In the absence of particle--particle collisions, it remains most likely that the scattering is caused by self-generated wave instabilities of the electron beam-plasma system \citep{Pagel-etal-2007, Bercic-etal-2019, Graham-etal-2021}.
These instabilities and their consequences have been studied to a certain extent, but with a system that is more simplified than in reality and considering only two components \citep{Saeed-etal-2017, Shaaban_mnras2018, Verscharen-etal-2019, Lee-etal-2019, Lopez-etal-2020, Innocenti-etal-2020, Schroder_2025ApJ}.

We report, for the first time (to our knowledge), a detailed modeling of these instabilities, as triggered, more realistically, by a three-component system of electrons: core, halo, and strahl.
Moreover, in Sect.~\ref{sec:model} the three components are realistically modeled by Maxwellian velocity distributions for the core, as well as non-Maxwellian, generalized Kappa ($\kappa$-)power laws for the halo and strahl populations \citep{Stverak-etal-2008, Wilson-etal-2019a, Wilson-etal-2019b, Scherer-etal-2022}.
Therefore, the wave dispersion and stability equations become more complicated, even for waves propagating parallel to the magnetic field (see Sect.~3), which we solved using advanced numerical solvers.
Modeling with standard Kappa distributions (SKDs) allows the derivation of the analytical expressions of these equations in terms of specific dispersion (integral) functions, which can be solved using the Dispersion Solver for Kappa Plasmas (DIS-K; made available by \cite{rodrigo_a_lopez_DISK}).
This applies to any plasma system with anisotropic Maxwellian and SKD populations \citep{Lopez_Shaaban_Lazar_2021, Lopez2021_book}.
{In contrast, modeling with newer regularized Kappa distributions (RKDs) makes the expressions of the dispersion equations and the related integral functions even more difficult to obtain analytically; currently only those reduced to parallel electrostatic wave modes are known \citep{Scherer2018, Gaelzer-Fichtner2024}.}
RKDs were introduced to fix the inconsistencies and limitations of SKDs, {in the sense that the RKD
ensures convergence of higher order velocity moments for all $\kappa >0$ while preserving the observed suprathermal behavior} \citep{Scherer2018, Scherer_2019a, Lazar-Fichtner-2021}. {The SKD is only well defined for $\kappa > 3/2$. This lower bound is mathematically imposed rather than physically motivated, and it therefore constitutes a formal restriction of the model, 
even though space plasmas with $\kappa < 3/2$ are observed} \citep{Gloeckler-etal-2012}\footnote{{The exploitation of RKDs should also motivate observational analysis \citep{Scherer-etal-2022}; fitting the VDs observed in situ with RKDs brings physical consistency to the parameterizations obtained even for moderate suprathermal tails, such as those described by a moderate but sufficiently small parameter of $\kappa > 3/2$, because RKDs significantly reduce the proportion and effect of unphysical superluminal particles (with speeds exceeding speed of light in a vacuum) from the SKD tails \citep{Scherer_2019a}.}}. {Therefore, using the RKD offers improved mathematical and physical consistency and allows self-consistent moment closures.}
In this case, one can still operate with general dispersion equations into which the numerically discretized VDs are introduced in order to solve them with the most advanced dedicated solvers with arbitrary VDs \citep{Husidic-etal-ASS-2021}.
Here, we exploited the Arbitrary Linear Plasma Solver (ALPS; \citep{Verscharen_2018, ALPS:2023}, which has already been tested for parallel waves and instabilities \citep{Verscharen_2018, SchroderPoP2025, Schroder_2025ApJ, Tischmann-etal-2025}, particularly in the recent analysis of heat-flux instabilities with a simplified model of only two electron core and strahl populations \citep{Schroder_2025ApJ}.

This primary work focused on unstable electromagnetic solutions describing heat-flux instabilities by minimizing the effects of possible (intrinsic) temperature anisotropies.
Even so, the results, as shown in Sect.~4, are markedly different from those previously obtained with only two (core-strahl) components. 
The growth rates depend on the (relative) densities and thermal or suprathermal spreads of these components, indicating an interplay between two unstable modes triggered by relative beam speeds, the core-beam and the halo-beam drifts.
Whistler heat-flux instabilities (WHFIs) with right-handed (RH) polarization can thus be excited, as well as the fire-hose heat-flux instabilities (FHFIs) from the lower frequency branch with left-handed (LH) polarization \citep{Gary_1993, Shaaban_mnras2018, Lopez-etal-2019b, Innocenti-etal-2020}.
It is expected that the effects of the two excited modes can accumulate if they are of the same nature, i.e., both have RH polarization, otherwise they still remain in competition. {While whistler waves associated with electron heat flux 
are frequently observed in the SW and propagating mostly quasi-parallel to the magnetic field \citep{Lacombe-etal-2014, Tong_APJL_2019, Jagarlamudi-etal-2020, Cattell-etal-2020, Kretzschmar-etal-2021}, observational evidence of electron-driven fire-hose fluctuations is comparatively less abundant. 
Conclusive observations of electron-fire-hose fluctuations have only recently emerged, but outside the pristine SW, in the reconnection jets in the earth's magnetotail \citep{Cozzani_JGR2023, Zhang_GRL2025}.  
Detecting FHFI-amplified waves may be difficult because their favorable parameter regime is significantly narrower than that of the WHFI \citep{Shaaban_mnras2018, Lopez-etal-2020}.
Therefore, their elusive nature in the SW may stem from technical limitations rather than their absence, as simulations demonstrate the potential of FHFIs, particularly for the regulation of electron heat flux under the SW conditions \citep{Lopez_etal2019, Lazar-etal-2023b, Innocenti-etal-2020}. 
}
In Sect.~\ref{sec:conclusion} we draw the main conclusions on the importance of the new realistic modeling and discuss perspectives for further development and application.

\section{Three-component electron model based on observations} \label{sec:model}
%%%%%%%%%%%%%%%%%%%%%%%%%%%%%%%%%%%%

In the observed electron VDs we can identify the three components \citep{Pierrard-etal-2001, Wilson-etal-2019a, Wilson-etal-2019b, Abraham-etal-2022}, the core (subscript $c$), the halo (subscript $h$), and the strahl (subscript $s$):
\begin{equation}\label{e1}
f(v_{\parallel}, v_{\perp}) = \frac{n_c}{n} f_c(v_{\parallel}, v_{\perp}, u_c) +  \frac{n_h}{n} f_h(v_{\parallel}, v_{\perp}, u_h) +\frac{n_s}{n} f_s(v_{\parallel}, v_{\perp}, u_s).
\end{equation}
Here, $n = n_c + n_h + n_s$ denotes the total electron number density, and $n_i/n$ ($i=c, h, s$) are the relative densities. 
The variables $v_{\parallel}$ and $v_{\perp}$ refer to the velocity components parallel and perpendicular to the background magnetic field, respectively, and $u_i$ ($i=c, h, s$) are the relative drifts in the proton rest frame.
In general, the strahl exhibits a major antisunward drift (or relative velocity) of $u_s > 0$, while the high-density core exhibits a sunward drift of $u_c<0$.

The core is a drifting Maxwellian:
\begin{equation}\label{e2}
f_c(v_\parallel, v_\perp) = \frac{1}{\pi^{3/2} \theta_{c,\perp}^2 \theta_{c,\parallel}} \exp\left(-\frac{(v_\parallel - u_c)^2}{\theta_{c,\parallel}^2} - \frac{v_\perp^2}{\theta_{c,\perp}^2}\right),
\end{equation}
with thermal speeds of $\theta_{c \|, \perp} = \sqrt{2k_B T_{c,\|, \perp}/m}$, a particle mass of $m$, Boltzmann constant of $k_B$, and temperature of $T_{c,\|, \perp}$.
The halo ($j=h$) and the strahl ($j=s$) component are modeled with Kappa VDs, and we first considered a drifting SKD:
\begin{equation}\label{eq:fSKD}
f_{\text{SKD}}(v_\parallel, v_\perp,\kappa) = \frac{1}{\pi^{3/2} \theta_\perp^2 \theta_\parallel} \frac{\Gamma(\kappa + 1)}{\Gamma(\kappa - 1/2)} \left[1 + \frac{(v_\parallel - u_j)^2}{\kappa \theta_\parallel^2} + \frac{v_\perp^2}{\kappa \theta_\perp^2}\right]^{-\kappa -1}.
\end{equation}
$\Gamma$ represents the Gamma function, and the exponent $\kappa$ models the high-energy tails.
The temperatures are $\kappa$-dependent: 
\begin{align} \label{e4}
T_{\kappa, \|,\perp}= \dfrac{\kappa}{\kappa-3/2} \dfrac{m \theta_{\|,\perp}^2}{2 k_B}= \dfrac{\kappa}{\kappa-3/2}  T^{M}_{\|,\perp}>T^{M}_{\|,\perp}, 
\end{align}
and $T^{M}_{\|,\perp} = m \theta_{\|,\perp}^2 / (2 k_B) =\lim \limits_{\kappa \to \infty} T_{\kappa, \|,\perp}$ is the lowest temperature, at which the halo does not have suprathermal tails and can be reproduced by a drifting Maxwellian, such as \eqref{e2}.
In a {quasi-}neutral plasma, drift speeds satisfy the zero-current condition $n_c u_c + n_h u_h + n_s u_s = 0$. 
According to observations \citep{Stverak-etal-2008, Wilson-etal-2019a, Wilson-etal-2019b}, the core and halo may have comparable drift speeds, and for simplicity we assume them to be equal: $u_c = u_h = - n_su_s/(n_c+n_h)$. {By neglecting the core-halo drift, the present analysis may exclude the class of core–halo–driven
heat-flux instabilities. 
Of these, whistler heat flux instability has been extensively examined \citep{Saeed-etal-2016, Saeed-etal-2017, Shaaban_mnras2018, Shaaban_2019MNRAS, Vasko_POP2020, Lopez_etal2019, Micera-et-al-2020} and is supported by observational data \citep{Tong_APJL_2019, Tong_APJ_2019}. 
The present study instead isolated the role of instabilities driven by the major drift of the strahl component, decoding the interplay of two different excitations, generated not only by the core-strahl drift, which is already known, but also by the halo-strahl drift. }

\begin{figure*}[ht]\centering
\includegraphics[width=0.99\textwidth]{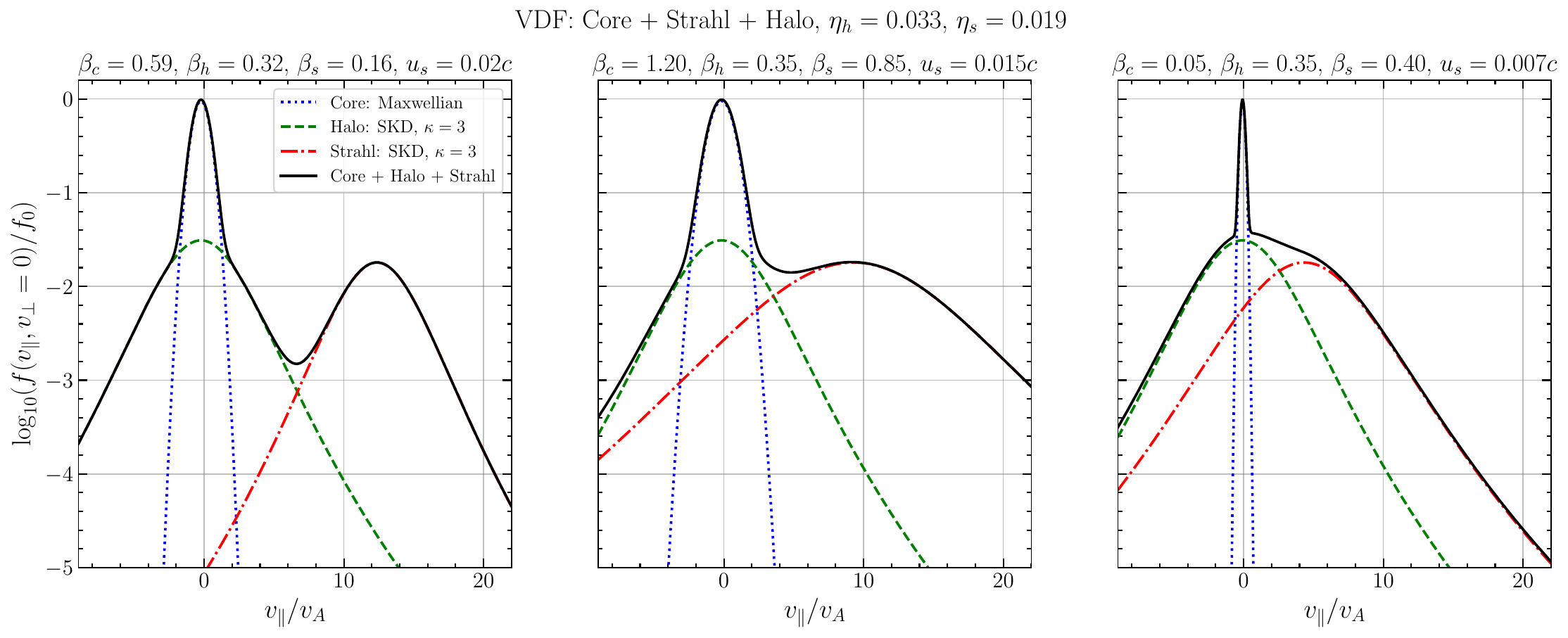}
\caption{Three-component VDs relevant for different unstable regimes: FHFI (left), WHFI + FHFI (middle), and WHFI (right). The Maxwellian core is plotted with blue dots, the halo (SKD) is shown with a dashed green line, the strahl (SKD) is shown with a dash-dotted red line, and the combination (total VD) of the two is shown with a solid black line.} \label{f1}
\end{figure*}

For the halo and strahl we also invoke more complex but also more consistent drifting RKDs:
\begin{equation}\label{eq: frkd}
\begin{aligned}
f_{\mathrm{RKD}}\left(v_{\|}, v_{\perp},\kappa, \alpha \right)= &  \frac{1}{\pi^{3 / 2} \theta_{\|} \theta_{\perp}^2 W} \left(1+\frac{(v_{\|}-u_s)^2}{\kappa \theta_{\|}^2}+\frac{v_{\perp}^2}{\kappa \theta_{\perp}^2}\right)^{-\kappa-1} \\ &\times \exp \left(-\frac{\alpha_{}^2 (v_{\|}-u_s)^2}{\theta_{\|}^2}-\frac{\alpha_{}^2 v_{\perp}^2}{\theta_{\perp}^2}\right),
\end{aligned}
\end{equation}
with 
\begin{equation}
    W = U\left(\frac{3}{2}, \frac{3-2\kappa}{2},\alpha^2 \kappa \right)
\end{equation}
and $U$ denoting the Tricomi function \citep{Scherer_2019}. The dimensionless cutoff parameter, $\alpha$, controls the exponential decay strength and does not depend on $\kappa$.
We demonstrate that such realistic three-component models, combining the electron core, halo, and strahl, can also consider RKDs, and resolving their dispersion and stability properties is still possible numerically.

In the present analysis we focus on heat-flux instabilities generated by the major drift of the electron strahl, in the excitation of which protons generally play a minimal role. 
Therefore, we reduced protons to a single component with a Maxwellian distribution as in Eq.~\eqref{e1}, but with the parameters identified by the subscript $p$, namely thermal speed $\theta_p = \sqrt{2k_B T_p/{m_p}}$, particle mass $m_p$, and temperature $T_p$. 

Figure~\ref{f1} shows three generic examples of such VDs (solid black line), combining three components: the core (dotted blue line), the halo (dashed green line), and the strahl (dash-dotted red line). 
We plot log$_{10}(f(v_\parallel, v_\perp=0) / f_0)$ as a function of (normalized) parallel velocity, $v_\parallel / v_A$, where $f_0 = f(v_\parallel=0, v_\perp=0 )$ and $v_A = B_0 / \sqrt{4\pi n_j m_j}$ is the Alfv\'en speed.
These VD models were chosen to be relevant for the three examples of unstable regimes studied in Sect. 4, namely when both growth rate peaks are of the same nature; i.e., both FHFIs (left panel) or both WHFIs (right panel), but also when these peaks are different in nature, combining both FHFI and WHFI (middle panel).
Indicated in the titles are the values used for the key parameters; these are the relative densities ($\eta_j= n_j/n_e$) for the halo ($j=h$) and strahl ($j=s$); the parallel plasma beta of each population, $\beta_j = 8 \pi n_j k_B T_{j \parallel} /B_0^2$, including the core ($j=c$); and the relative drift of the strahl ($u_s$, in units of speed of light in vacuum $c$).

In general the core is much cooler than the halo and the strahl.
The left panel in Fig.~\ref{f1} shows the most contrasting situation with a significant relative drift of the strahl, markedly exceeding thermal velocities of the halo and the strahl (which are comparable).
In the right panel, the relative drift speed of the strahl is significantly reduced and becomes comparable to its thermal velocity.
The suprathermal populations, in particular the strahl, must still be hotter than the core.
This configuration has an effective temperature anisotropy in the perpendicular direction, and it can typically trigger WHFI \citep{Shaaban_mnras2018, Lopez-etal-2020}.
In the middle panel we have a transition regime with a similar distribution to the one in the left panel, but in which the strahl has a {lower} drift and a slightly higher temperature than the halo.
This allowed us to obtain two peaks of different natures: an FHFI driven by the strahl-halo drift and a WHFI driven by the strahl-core drift.
In the present analysis, we start from these distinct configurations, varying the key parameters on which the dispersion and stability properties depend, in accordance with in situ observations. 
We {were} mainly inspired by data collected at 1~AU \citep{Stverak-etal-2008, Wilson-etal-2019a, Wilson-etal-2019b}, although data from lower heliocentric distances may also be relevant for heat-flux instabilities \citep{Stverak-etal-2008, Abraham-etal-2022}.
In order to isolate the effects of the drifts, in this first study we neglected temperature anisotropies, $A_{j} = T_{\perp,j}/T_{\|,j} = 1$, and, for all cases, we adopted a core-halo density contrast of $\eta_h = n_h/n_c = 0.033$ and a core-strahl density contrast of $\eta_s = n_s/n_c = 0.019$ with the normalized densities $n_c/n_0 = 0.951$, $n_h/n_0 = 0.031$, $n_s/n_0 = 0.018$. 
The chosen electron plasma-to-gyrofrequency ratio is $\omega_{p,e}/|\Omega_e| = 619$. 
The core is generally (quasi-)Maxwellian \citep{Stverak-etal-2008}, and even when modeled with Kappa VDs, sufficiently large values of $\kappa > 6$ are reported \citep{Wilson-etal-2019a, Wilson-etal-2019b} that justify the choice of $f_c=f_M$.

%%%%%%%%%%%%%%%%%%%%%%%%%%%%%%%
\section{Dispersion formalism }
%%%%%%%%%%%%%%%%%%%%%%%%%%%%%%%

We first analyze the distributions described by "classical" models such as Maxwellian and SKD ones, whose dispersion and stability properties can be reliably solved by the DIS-K solver \citep{Lopez_Shaaban_Lazar_2021, rodrigo_a_lopez_DISK}.
The solutions thus obtained allowed us to test the ALPS solver and show that it can be exploited for the direct solution of such three-component VDs described by RKDs, without resorting to the related dispersion equations whose derivation is otherwise very complicated.

The linearized general dispersion relation for the parallel propagation ($\Vec{k} \times \Vec{B_0 }=\Vec{0}$, where $\Vec{k}$ is the wave vector) can be written as \citep{Lazar_JGR_2014, Shaaban_mnras2018}
\begin{equation}
\begin{aligned}
\frac{k^2 c^2}{\omega^2}=& \ 1  +\frac{4 \pi^2}{\omega^2} \sum_{j=c, h, s, p} \frac{q_j}{m_j} \int_{-\infty}^{\infty} \frac{d v_{\|}}{\omega-k v_{\|} \pm \Omega_j} \\ &\times \int_0^{\infty} d v_{\perp} v_{\perp}^2 
 \times\left[\left(\omega-k v_{\|}\right) \frac{\partial f_j}{\partial v_{\perp}}+k v_{\perp} \frac{\partial f_j}{\partial v_{\|}}\right],
\end{aligned}
\end{equation}
with
the complex $\omega(k) = \omega_r (k) + i\gamma(k)$ as a function of wave number, $k$, in terms of the wave frequency,  $\omega_r=\Re(\omega)$, and the growth or damping rate, $\gamma=\Im(\omega)$.
$f_j$ is the VD of the $j$-th population, protons ($j=p$); and electron core ($j=c$), halo ($j=h$), and strahl ($j=s$) components. 
For our model with three-component electron VDs and one proton population (under the assumption of no temperature anisotropy), the dispersion relation becomes
\begin{equation}\label{e8}
\begin{aligned}
\left(\frac{kc}{\omega_{p,e}} \right)^2 =\  &\frac{\omega/ |\Omega_e|}{(kc/\omega_{p,e}) \sqrt{\mu \beta_p}} Z_{M}\left(\frac{\mu \omega/ |\Omega_e| \pm 1}{(kc/\omega_{p,e}) \sqrt{\mu \beta_p}}\right)\\
&+\frac{n_c}{n_0} \frac{\left(\omega/ |\Omega_e|-U_c (kc/\omega_{p,e})\right)}{(kc/\omega_{p,e}) \sqrt{\beta_c}}\\
&\times Z_{M}\left(\frac{\omega/ |\Omega_e| \mp 1-U_c (kc/\omega_{p,e})}{(kc/\omega_{p,e}) \sqrt{\beta_c}}\right) \\
& +\frac{n_h}{n_0}  \frac{\left(\omega/ |\Omega_e|-U_h (kc/\omega_{p,e})\right)}{(kc/\omega_{p,e}) \sqrt{\beta_h}} \\
&\times Z_{\kappa }\left(\frac{\omega/ |\Omega_e| \mp 1-U_h (kc/\omega_{p,e})}{(kc/\omega_{p,e}) \sqrt{\beta_h}}\right) \\
& +\frac{n_s}{n_0}  \frac{\left(\omega/ |\Omega_e|-U_s (kc/\omega_{p,e})\right)}{(kc/\omega_{p,e}) \sqrt{\beta_s}} \\
&\times Z_{\kappa }\left(\frac{\omega/ |\Omega_e| \mp 1-U_s (kc/\omega_{p,e})}{(kc/\omega_{p,e}) \sqrt{\beta_s}}\right),
\end{aligned}
\end{equation}
using $U_j=u_j \omega_{pe}/(c |\Omega_e|)$ with $j=c,h,s$, and $U_p=0$.
The plasma frequency of species $j$ is defined as $\omega_{p,j} = \sqrt{4 \pi n_j q_j^2 / m_j}$, and the corresponding nonrelativistic gyrofrequency as $\Omega_j = q_j B_0 / (m_j c)$, where $m_j$, $q_j$, and $n_j$ denote the rest mass, electric charge, and number density of species $j$, protons ($j = p$), and electrons ($j = e$), respectively. 
The parallel plasma beta parameters are defined as $\beta_{j} = 8 \pi n_j k_B T_{j,\|} / B_0^2$ for protons ($j=p$) and each electron component, core ($j=c$), halo ($j=h$), and strahl ($j=s$). 
We adopted a realistic proton-to-electron mass ratio of $\mu = m_p/m_e = 1836$, while the parallel thermal velocity of the $j$-th electron component is expressed as $\theta_{j, \|}/c = \sqrt{\beta_j}\, |\Omega_e| / \omega_{p,e}$. 
{In Eq. (\ref{e8})} 
\begin{equation}
Z_{M}\left(\xi_{M, j}^{ \pm}\right)=\frac{1}{\pi^{1 / 2}} \int_{-\infty}^{\infty} \frac{\exp \left(-x^2\right)}{x-\xi_{M, j}^{ \pm}} d x, \quad \mathfrak{J}\left(\xi_{M, j}^{ \pm}\right)>0
\end{equation}
is the plasma dispersion function for Maxwellian populations \citep{Fried-Conte-1961}, with {the argument $\xi_{M, j}^{ \pm}=(\omega\pm |\Omega _j| - k u_j)/(k\theta_{j ,\|})$, where $u_p =0$,} and $\pm$ differentiates between right-handed (RH) and left-handed (LH) circular polarization; i.e., electron arguments, $\xi_{M, e}^{-}$, and the proton argument, $\xi_{M, p}^{+}$, describe RH-modes, and vice versa for LH-modes. 

\begin{equation}
\begin{aligned}
Z_{\kappa}\left(\xi_{\kappa, j}^{ \pm}\right)  =& \frac{1}{\pi^{1 / 2} \kappa^{1 / 2}} \frac{\Gamma(\kappa)}{\Gamma(\kappa-1 / 2)} \\ &\times \int_{-\infty}^{\infty} \frac{\left(1+x^2 / \kappa\right)^{-\kappa}}{x-\xi_{\kappa,j}^{ \pm}} d x, \quad \mathfrak{J}\left(\xi_{\kappa, j}^{ \pm}\right)>0
\end{aligned}
\end{equation}
is the (modified) plasma dispersion function for SKD populations  \citep{Hellberg-Mace-2002, Lazar-etal-2008}, with the argument $\xi_{\kappa, j}^{ \pm}=(\omega\pm |\Omega _e| - k u_j)/(k\theta_{j,\|})$, ($j= h, s$). {Although the dispersion relation formally retains the same tensor structure as in the Maxwellian or SDK cases, the evaluation of the dielectric response for a drifting regularized Kappa (RKD) distribution requires modification of the velocity integrals and the associated plasma dispersion functions, which generally cannot be expressed in a straightforward manner in terms of standard functions, resulting in technical complexity \citep{Gaelzer-Fichtner2024}. Therefore,
deriving the dispersion formalism for anisotropic (particularly drifting) RKDs remains nontrivial}, and we employed ALPS for these cases, numerically solving the most general dispersion relation for arbitrary shapes of the VDs in hot plasmas; see Appendix~\ref{appendixA}. 
To initialise ALPS, we provide numerical values of the VD $f_{0j} (p_{\|},p_{\perp})$ on a momentum space grid where $p_{\|, \perp}$ is the parallel and perpendicular momentum with respect to the background field. This corresponds to discretising Eq.~\eqref{f1} into a structured ASCII table. 
Apart from numerical control parameters, the only additional input required is an initial guess for $\omega_r$ and $\gamma$. 

{Dealing with new properties of more complex three-component electron models, in this first analysis we limited ourselves to waves propagating parallel to the magnetic field. 
This limitation is somewhat justified at least for the whistler heat-flux waves observed in solar wind, whose spectra appear dominated by quasi-parallel propagating waves \citep{Lacombe-etal-2014, Tong_APJL_2019, Jagarlamudi-etal-2020, Kretzschmar-etal-2021}.
Although oblique whistler modes are also reported \citep{Cattell-etal-2020}.
However, it should be noted that theoretical and numerical approaches predict oblique whistlers to be more effective in pitch-angle scattering and heat-flux regulation under typical solar wind conditions \citep{Micera-et-al-2020, Kuzichev_2019, Kuzichev_2023}.
Further refinement of theoretical and numerical models could therefore be essential to resolving this inconsistency.}
%
%%%%%%%%%%%%%%%%%%%%%%%%%%
\section{Unstable heat-flux solutions}
%%%%%%%%%%%%%%%%%%%%%%%%%%

%%%
\subsection{Two peaks of FHFIs}
%%%

\begin{figure*}[t!]\centering
\includegraphics[width=0.9\textwidth]{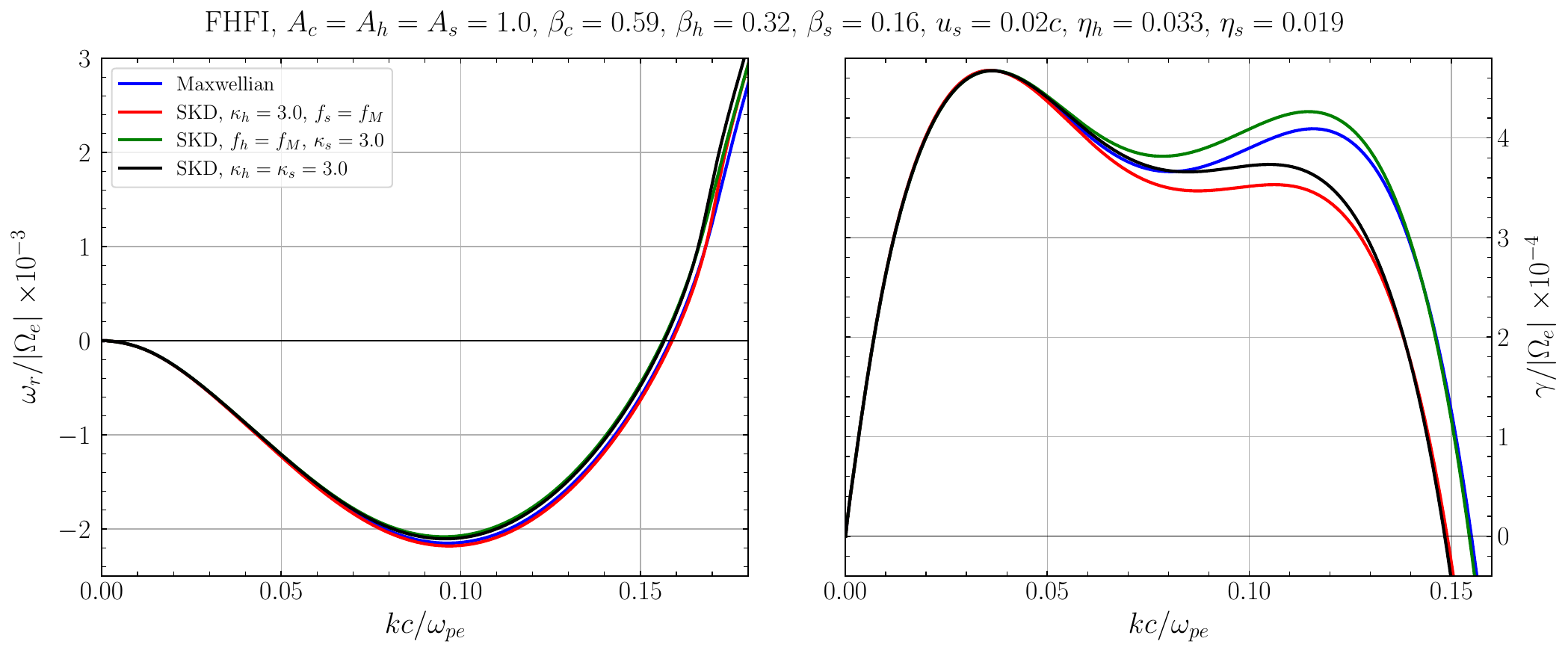}
\caption{FHFI solutions derived with DIS-K for a Maxwellian core and different halo-strahl combinations (legend): Maxwellian halo and strahl (blue), SKD halo and Maxwellian strahl (red), Maxwellian halo and SKD strahl (green), and SKD halo and strahl (black). 
The growth rates, $\gamma/|\Omega_e|$ (right), show two (comparable) peaks of instabilities with the same LH polarization indicated by the sign of $\omega_r/|\Omega_e| < 0$ (left). The parameters used are indicated in the title.}\label{f2}
\end{figure*}

We first employed DIS-K to solve the dispersion Eq. \eqref{e8}, and the results are presented in Figs.~\ref{f2} and \ref{f4}, showing both the real (normalized) frequency, $\omega_{r}/|\Omega_{e}|$ (left panels), and the corresponding (normalized) growth rates, $\gamma/|\Omega_{e}|$ (right panels). 
This solver has been tested for various VD configurations including those relevant for HFIs \citep{Lopez_Shaaban_Lazar_2021,Lopez2021_book,rodrigo_a_lopez_DISK, Schroder_2025ApJ}.  

\begin{figure*}[th!]\centering
\includegraphics[width=0.99\textwidth]{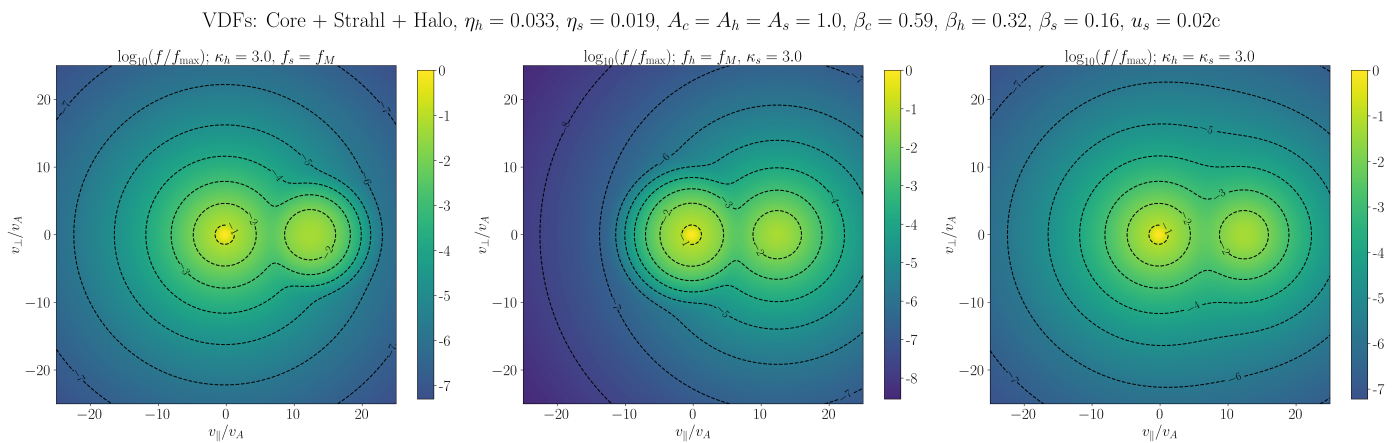}
\caption{Contour plots of the normalized electron VDs, $\log(f/f_{\text{max}})$, in the $(v_\parallel,v_\perp)$-plane for different SKD configurations relevant to the solutions in Fig.~\ref{f2}, illustrating the drift-induced anisotropies and suprathermal tails. Left: Broadened core halo ($\kappa_h=3.0$) and a sharp decay in the antisunward direction due to the Maxwellian strahl. Middle: More compact core halo with a steep Maxwellian decay of the halo and a slower decay of the SKD strahl. Right: Both halo and strahl have SKDs, yielding a global topology dominated by the denser halo and explaining the unstable solutions. The parameters used are indicated in the title.
} \label{f3}
\end{figure*}  

To explore various contributions of SKDs, in Fig.~\ref{f2} four representative cases are compared: a purely Maxwellian case when all core, halo, and strahl components are Maxwellian, $f_{c,h,s} \to f_M$ (blue), and three suprathermal configurations, with an SKD halo ($\kappa_{h}=3.0$), $f_h \to f_{\rm SKD}$, while keeping $f_s \to f_M$ (red), with an SKD strahl ($\kappa_{s}=3.0$), keeping $f_h \to f_M$ (green), and when both the halo and strahl have SKDs ($\kappa_h = \kappa_s = 3.0$, black). 
Growth rates exhibit two peaks of FHFI in all cases: a low-wave-number LH mode triggered by the core-strahl drift and not significantly affected by suprathermal populations in SKD models, and a high-wave-number LH mode driven by the halo-strahl drift and highly sensitive to the suprathermal components. 
Previous studies have only analyzed the first (core-strahl FHFI) in similar intervals of unstable wave numbers \citep{Shaaban_mnras2018, Shaaban_2018PoP}, which helped us to identify the nature of the corresponding peak. 
In addition we find that this can be in competition with the new unstable halo-strahl FHFI mode with comparable growth rates but {higher} wave numbers.
For the chosen parameters, the Maxwellian reference case (blue) exhibits a slightly dominant FHFI peak at $kc/\omega_{\mathrm{pe}}\approx 0.036$ with a maximum growth rate of $\gamma_{\text{max}}/|\Omega_{e}| \approx 4.68 \times 10^{-4}$, and a second FHFI branch at $kc/\omega_{\mathrm{pe}} \approx 0.12$ with a maximum growth rate of $\gamma_{\text{max}}/|\Omega_{e}| \approx 4.09 \times 10^{-4}$. 
Introducing a suprathermal SKD tail in the halo (red), $\kappa_{ h}=3.0$, decreases the second peak (compared to the Maxwellian case) by a factor of $\sim 0.86$ to $\gamma_{\text{max}}/|\Omega_{e}| \approx 3.53 \times 10^{-4}$ and shifts it to lower wave numbers of $kc/\omega_{\mathrm{pe}} \approx 0.106$ (with an overall decrease of the range of unstable wave numbers). 
The halo is denser than the strahl, and increasing its tails also reduces the effective anisotropy given by the relative halo-strahl drift, leading to a decrease of the second peak to  $\gamma_{\text{max}}/|\Omega_{e}| \approx 3.73 \times 10^{-4}$.
Modeling only the strahl population with an SKD (green, $\kappa_s=3.0$) leads to the opposite effect, thus increasing the maximum growth rate by a factor of $1.04$ to $\gamma_{\text{max}}/|\Omega_{e}| \approx 4.27 \times 10^{-4}$ at $kc/\omega_{\mathrm{pe}} \approx 0.115$.
In this case the strahl becomes hotter than the halo, and we may have a combined effect of FHFI and WHFI, because $\omega_r \to 0$ tends to change sign, and the latter is typically stimulated by suprathermal tails in the strahl component.
Otherwise, this can also be an effect associated with so-called fire-hose beaming instabilities driven by two counterbeaming populations (halo and strahl) with similar properties \citep{Moya_2022, Lazar-etal-2023b}, which still need to be examined for SKD electrons. 
In the last case, when both halo and strahl populations have SKDs (black), a cumulative effect is obtained, including the inhibiting effect of the halo on the second peak.

These differences can be correlated with the contrasts in the two-dimensional {VDs} in Fig~\ref{f3}, which shows contour plots of $\log(f/f_{\text{max}})$ in the $(v_\parallel,v_\perp)$-plane for the three different cases with SKDs. 
The first panel (left, $\kappa_h=3.0$) is relevant for the red solution in Fig.~\ref{f2}.
It depicts a broadened core-halo distribution, contrasting with the sharp decrease of the Maxwellian strahl in the antisunward direction.
Conversely, the middle panel exhibits a slower decay in the antisunward direction due to the SKD strahl ($\kappa_s=3.0$), but a more restrained core-halo distribution with a steep, contrasting gradient in the sunward direction that can eventually be determined as responsible for the highest peak (green) in Fig.~\ref{f2}. 
When both the halo and strahl have SKDs (right,  $\kappa_{h,s}=3.0$), the gradients are less steep in both directions, and the halo-strahl peak is again reduced (black solution in Fig.~\ref{f2}).     

\begin{figure*}[t!]\centering
\includegraphics[width=0.9\textwidth]{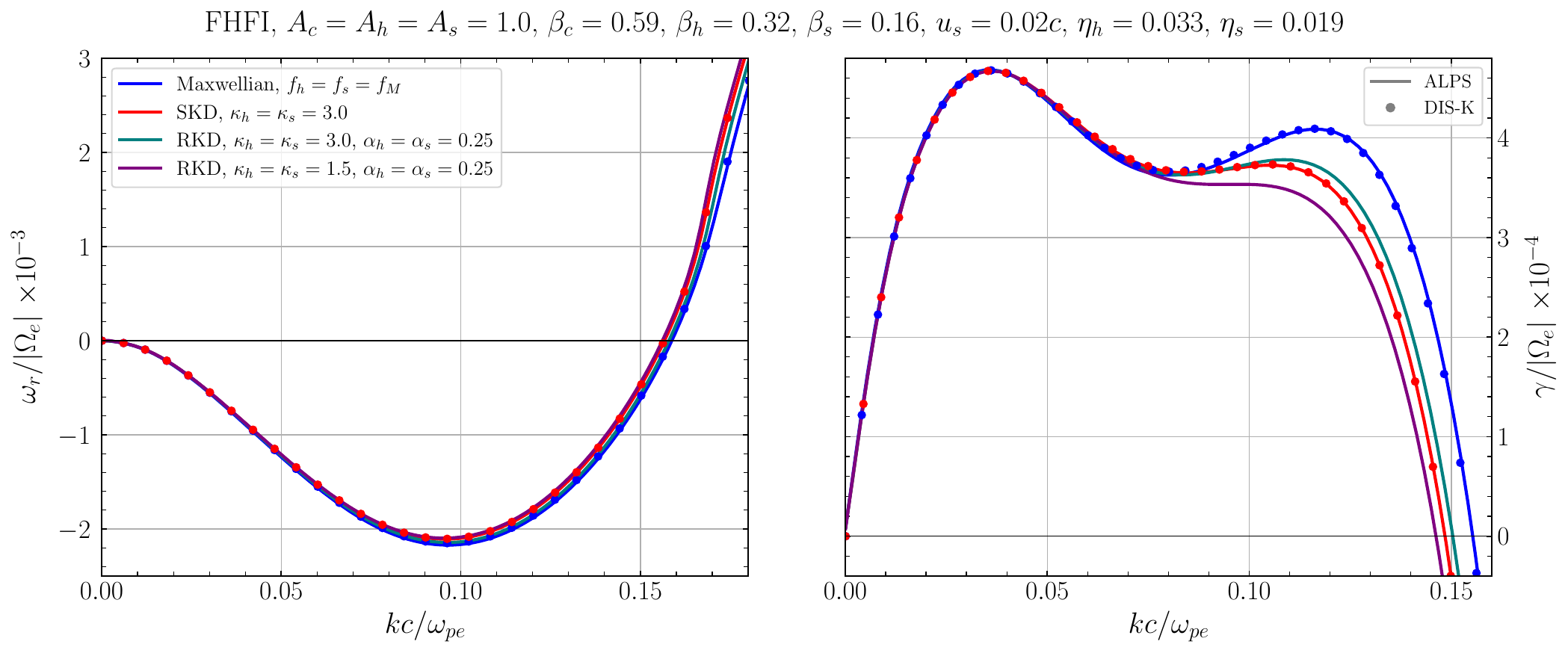}
\caption{Comparison of FHFI solutions, frequencies $\omega_r/ |\Omega_e|$ (left), and growth rates $\gamma/|\Omega_e|$ (right) as functions of wave number, $kc/\omega_{pe}$, obtained with DIS-K (dots) and ALPS (solid lines) for the reference case in Fig.~\ref{f2}, including the halo and strahl with Maxwellian (blue) and SKDs (red) and two additional configurations with the RKD halo and strahl populations ($\alpha_h = \alpha_s=0.25$): $\kappa_h=\kappa_s=3.0$ (green) and $\kappa_h=\kappa_s=1.5$ (purple). 
The parameters used are indicated in the title. }\label{f4}
\end{figure*}

We extended these cases and compared them with the solutions obtained for the RKD halo and/or strahl populations with ALPS \citep{ALPS:2023, Verscharen-etal-2019}. 
Figure~\ref{f4} compares the results from DIS-K (dots) and ALPS (solid lines) for the case shown in Fig.~\ref{f2}, together with two additional cases employing an RKD with an isotropic cutoff of $\alpha_h = \alpha_s = 0.25$,  $\kappa_h = \kappa_s = 3.0$ (green), and $\kappa_h = \kappa_s = 1.5$ (purple).
The solutions in the last case are quite special, since $\kappa_h = \kappa_s = 1.5$ are forbidden for SKDs.
The excellent agreement between the solutions validates the applicability of the ALPS solver.
An RKD with the same $\kappa_h = \kappa_s = 3.0$ as the SKD and a moderate cutoff is more Maxwellian, and it reduces the effect of the suprathermals. 
Thus, while the first remains unaffected, the second peak is enhanced, shifting toward the Maxwellian solution. 
Introducing the RKD with lower $\kappa$ and the same moderate cutoff leaves the first FHFI peak essentially unchanged, while the second peak becomes weaker compared to the SKD case with $\kappa_h=\kappa_s=3.0$.

\begin{figure*}[t!] \centering
\includegraphics[width=0.9\textwidth]{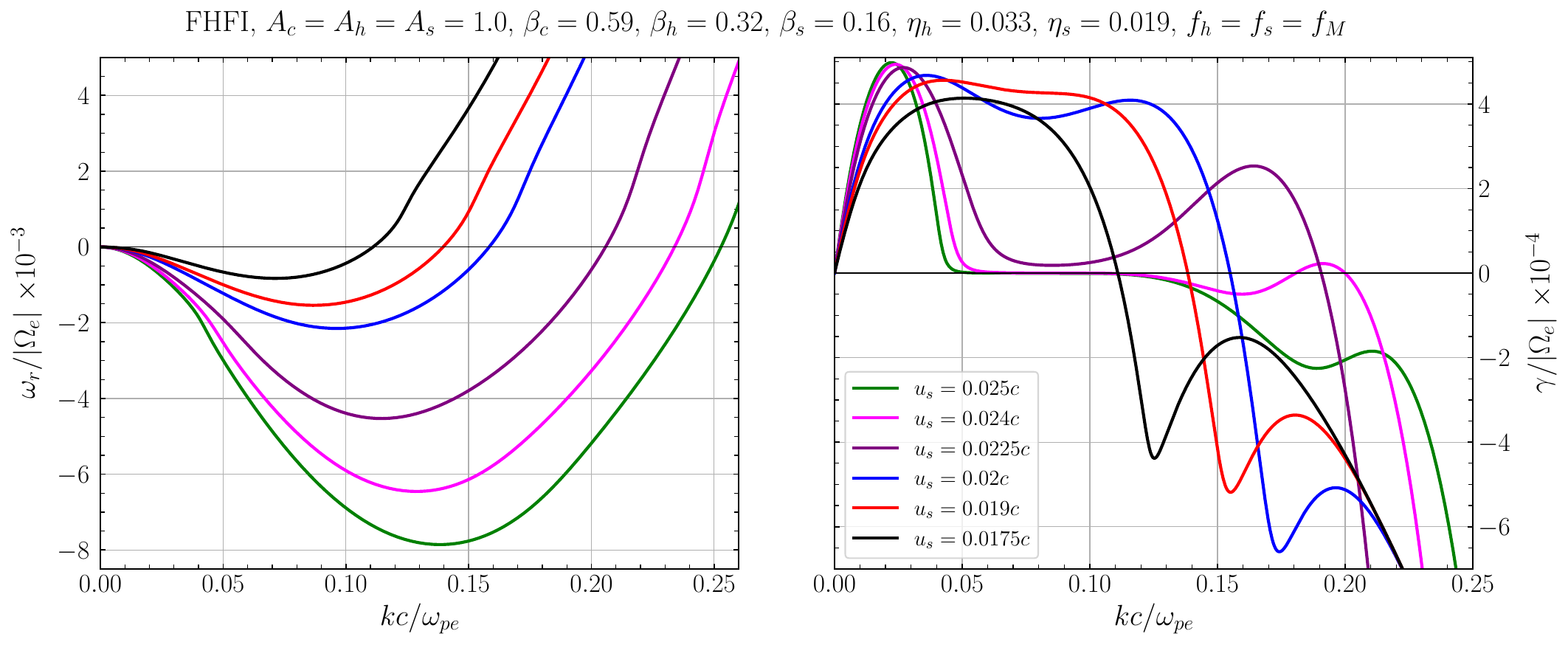} \caption{FHFIs in terms of frequency, $\omega_r/|\Omega_e|$ (left), and growth rate, $\gamma/|\Omega_e|$ (right), as functions of wave number $kc/\omega_{\mathrm{pe}}$, obtained with DIS-K for Maxwellian electron populations and different strahl speeds, $u_s$. 
The reference case ($u_s = 0.02c$) is compared with unstable solutions derived for lower drifts down to $0.0175c$ and also higher ones up to $0.025c$. 
The other parameters used are indicated in the title.} \label{f5}
\end{figure*}

To further study the driver of the FHFI, we examine the sensitivity of the instability to the relative drift velocity of the strahl. 
Figure \ref{f5} illustrates the variation of the real frequency, $\omega_r/|\Omega_e|$ (left panel), and the growth rate, $\gamma/|\Omega_e|$ (right panel), as a function of the normalized wave number $kc/\omega_{pe}$ for different strahl drift velocities $u_s$, ranging from $0.0175c$ to $0.025c$.
For simplicity, we assume all electron populations with Maxwellian VDs.
The case for $u_s=0.02c$ (blue) is already analyzed above and can be considered as the reference case. 
When the drift speed is reduced by $5\%$ to $u_s = 0.019c$, the maximum growth rate of the first peak is slightly weakened to $\gamma_{\text{max}}/|\Omega_{e}| \approx 4.56 \times 10^{-4}$ and shifted toward {higher} wave numbers: $kc/\omega_{\mathrm{pe}}\approx 0.042$.
The second peak, however, is shifted toward lower wave numbers: $kc/\omega_{\mathrm{pe}}\approx 0.075$.
Consequently, the two peaks approach each other and eventually overlap ($u_s = 0.0175c$), making the individual modes less distinguishable and 
yielding a single broad maximum with a (slightly) reduced growth rate of 
$\gamma_{\text{max}}/|\Omega_{e}| \approx 4.14 \times 10^{-4}$ around 
$kc/\omega_{\mathrm{pe}} \approx 0.051$. 
The overall range of unstable wave numbers, compared to the reference case, is reduced by a factor of $0.72$. Starting from the reference case, increasing the drift speed by $12.5\%$ to $u_s = 0.0225c$ leads to the opposite behavior: the two peaks separate. 
The first peak becomes sharper, shifts to lower wave numbers, and reaches an increased maximum growth rate of 
$\gamma_{\max}/|\Omega_{e}| \approx 4.87\times 10^{-4}$
at $kc/\omega_{\mathrm{pe}} \approx 0.027$. 
The second peak is shifted toward higher wave numbers and reduced by $37.9 \%$ to $\gamma_{\max}/ |\Omega_{e}| \approx 2.54\times 10^{-4}$ at $kc/\omega_{\mathrm{pe}} \approx 0.164$. 
A further increase to $u_s = 0.024c$ enhances both trends: the first peak becomes sharper and shifts to lower wave numbers, reaching $\gamma_{\max}/ |\Omega_{e}| \approx 4.94 \times 10^{-4}$ at $kc/\omega_{\mathrm{pe}} \approx 0.024$.
The second peak diminishes and moves to higher wave numbers, with $\gamma_{\max}/|\Omega_{e}| \approx 2.26 \times 10^{-5}$, which is more than an order of a magnitude smaller, at $kc/\omega_{\mathrm{pe}} \approx 0.191$. 
For higher drift speeds ($u_s = 0.025c$), the second mode becomes damped, leaving only a single sharp FHFI peak at low wave numbers with $\gamma_{\max}/|\Omega_{e}| \approx 4.98\times 10^{-4}$ at $kc/\omega_{\mathrm{pe}} \approx 0.022$.
These results delimit the range of relative drift velocities, $u_s$, for which two distinct FHFI peaks can be obtained for this specific set of plasma parameters.
%   

%%%%%%%%%%%%%%%%%%%%%%%%%%%%

\begin{figure*}[t]\centering
\includegraphics[width=0.9\textwidth]{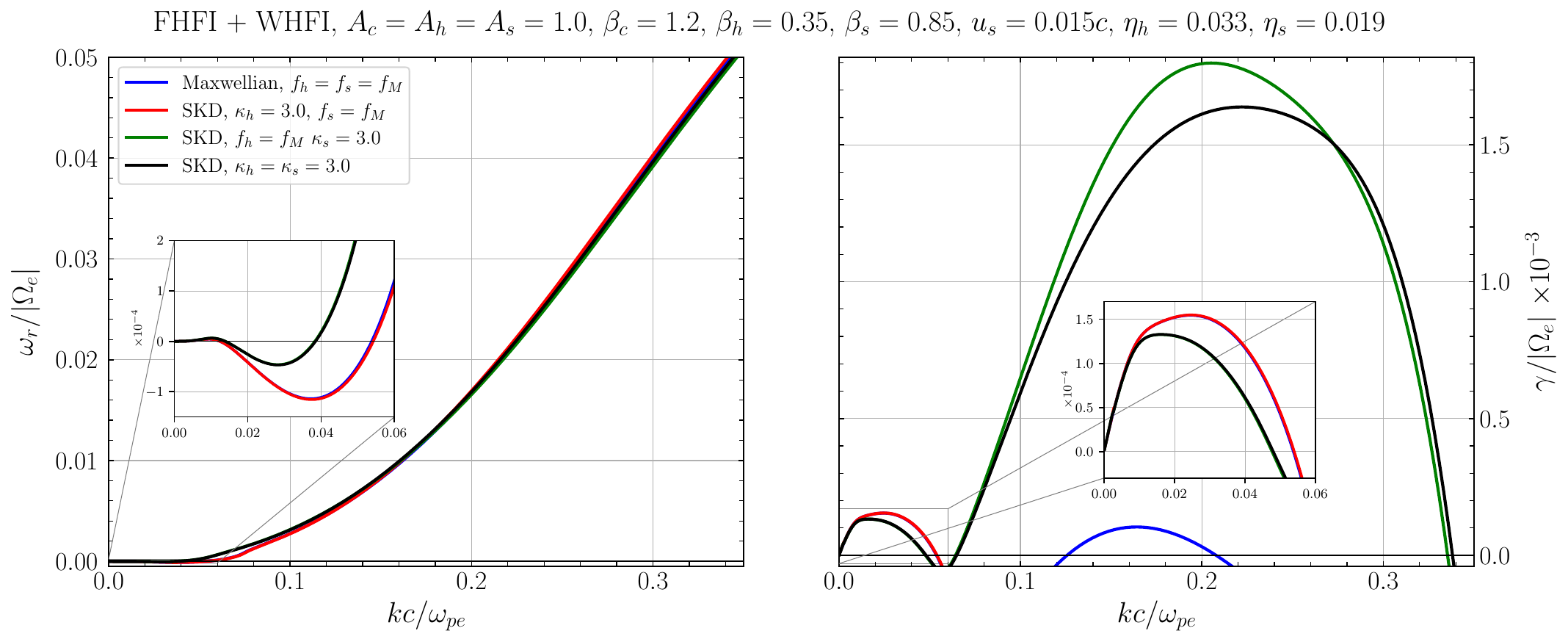}
\caption{Unstable solutions, $\omega_r/|\Omega_e|$ (left) and $\gamma/|\Omega_e|$ (right), with alternative peaks of FHFI and WHFI derived with DIS-K for a Maxwellian core and different halo-strahl combinations, respectively: Maxwellian--Maxwellian (blue), SKD--Maxwellian (red), Maxwellian--SKD (green), and SKD--SKD (black). The zoomed-in plot in left panel shows the change of sign and polarization corresponding to the low-wave-number peaks in the right panel. The parameters used are indicated in the legend and title.} \label{f6}
\end{figure*}

%%%
\subsection{WHFI and FHFI}
%%%
To obtain two peaks of instabilities of different natures, we adopted the plasma parameters in Fig.~\ref{f1}, middle panel.
The unstable solutions in terms of wave frequency (left) and growth rates (right) are shown in Fig.~\ref{f6} for  configurations comparing (as in Fig.~\ref{f2}) the reference case of Maxwellian electron populations (blue) with the other cases involving SKD halo and strahl components.
In the reference case, a first LH-polarized peak corresponding to an FHFI with $\gamma_{\max}/ |\Omega_{e}| \approx 1.55\times 10^{-4}$ appears at $kc/\omega_{\mathrm{pe}} \approx 0.025$. 
At higher wave numbers, the real frequency changes sign and a second, slightly broader peak with $\gamma_{\max}/|\Omega_{e}| \approx 1.04\times 10^{-4}$ arises at $kc/\omega_{\mathrm{pe}} \approx 0.164$; it is associated with an RH-polarized WHFI. 
We must remember that such transient regimes with two growth-rate peaks of different natures --respectively FHFI and WHFI-- were also obtained with models with only two electron populations, for example core and beam models \citep{Shaaban_mnras2018}.

For an SKD halo with $\kappa_h = 3.0$, the FHFI growth rate remains essentially unchanged, as proof that this instability is primarily driven by the core-strahl drift. 
The WHFI becomes stable, as evidence that this mode is triggered by the relative halo-strahl drift and is sensitive to the temperature contrast between these two populations. 
When the strahl has an SKD ($\kappa_s = 3.0$), the FHFI growth rate decreases slightly by $\sim 14.2\%$ and shifts to lower wave numbers, reaching $\gamma_{\max}/|\Omega_e| \approx 1.33\times 10^{-4}$ at $kc/\omega_{\mathrm{pe}} \approx 0.016$.
In contrast, the WHFI becomes visibly enhanced; a suprathermal and thus hotter strahl favors the WHFI (while the suprathermal halo suppresses it under the chosen conditions).
The peak is significantly broader and higher, with an increase of more than an order of magnitude to $\gamma_{\max}/|\Omega_e| \approx 1.80\times 10^{-3}$ at a higher $kc/\omega_{\mathrm{pe}} \approx 0.205$. 
The overall unstable wave-number range expands by a factor of three compared to the fully Maxwellian case, with $\gamma > 0$ occurring over both lower and higher wave numbers.     
When both halo and strahl have SKDs ($\kappa_h = \kappa_s = 3.0$), the FHFI peak remains essentially unchanged compared to the previous case where only the strahl followed an SKD, confirming that core-strahl drift is the dominant driver of the mode. 
The WHFI, however, exhibits a cumulative effect. 
Its peak is enhanced compared to the reference case due to the hotter strahl, but less than in the case when only the strahl has an SKD, because the temperature ratio between halo and strahl is now reduced, leading to $\gamma_{\max}/|\Omega_e| \approx 1.64\times 10^{-3}$ at $kc/ \omega_{\mathrm{pe}} \approx 0.222$. 
The range of unstable wave numbers remains comparable.

\begin{figure*}[ht]\centering
   \includegraphics[width=0.99\textwidth]{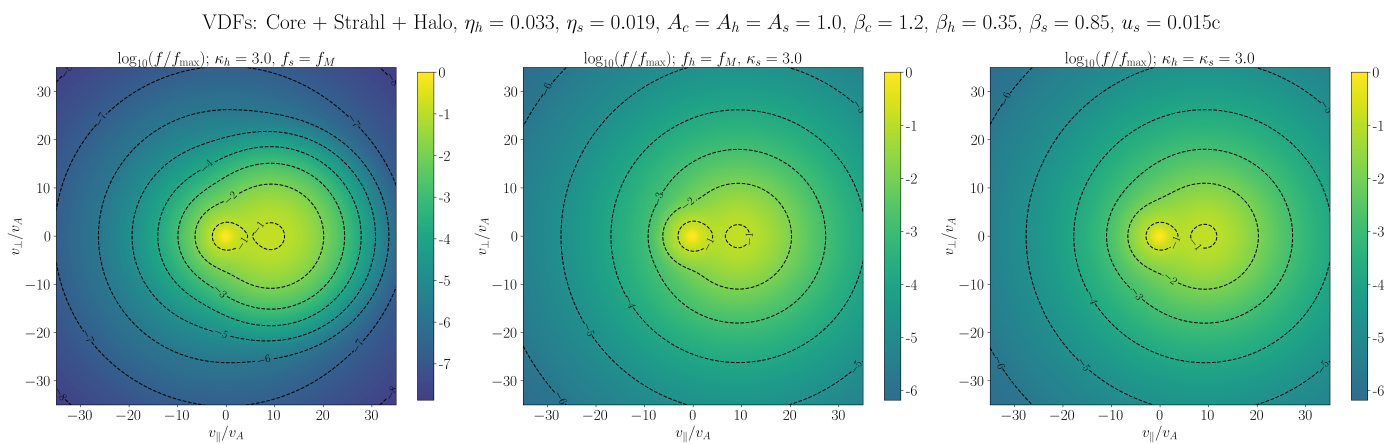}
\caption{Contour plots of the normalized electron VDs, $\log(f/f_{\text{max}})$, in the $(v_\parallel,v_\perp)$-plane with different SKD configurations relevant for the solutions in Fig.~\ref{f6}, illustrate the drift-induced anisotropies and suprathermal tails. 
Left: SKD halo ($\kappa_h = 3.0$) and a steeper falloff of the Maxwellian strahl in the antisunward direction. Middle: Maxwellian halo and a (much) hotter SKD strahl. Right: both the halo and the strahl have SKDs, and the topology is dominated by the hotter strahl (resembling the middle case) and is consistent with the corresponding growth rates. The other parameters used are indicated in the title.}\label{f7}
\end{figure*} 

These trends can again be understood from the corresponding VD contour plots shown in Fig.~\ref{f7}, which shows the underlying anisotropies. 
In the left panel, where only the halo has an SKD ($\kappa_h = 3.0$), the distribution exhibits a noticeably steeper decline compared to the other panels, especially in the antisunward direction due to the Maxwellian strahl. 
Instead, in the middle panel the strahl has an SKD ($\kappa_s = 3.0$), while the halo is Maxwellian. 
However, the influence of the less dense but much hotter strahl is already noticeable. 
In the right panel, with both populations following SKDs, the overall topology is again markedly dominated by the hotter strahl and therefore closely resembles the middle case.
These VDs are consistent with the behavior of the WHFI growth rates, also contrasting with the FHFI, where the cumulative effect was governed primarily by a denser and hotter halo.

%%%%%%%%%%%%%%%%%%%%%%
\subsection{Two peaks of WHFIs}
%%%%%%%%%%%%%%%%%%%%%%

\begin{figure*}[ht]\centering
\includegraphics[width=0.9\textwidth]{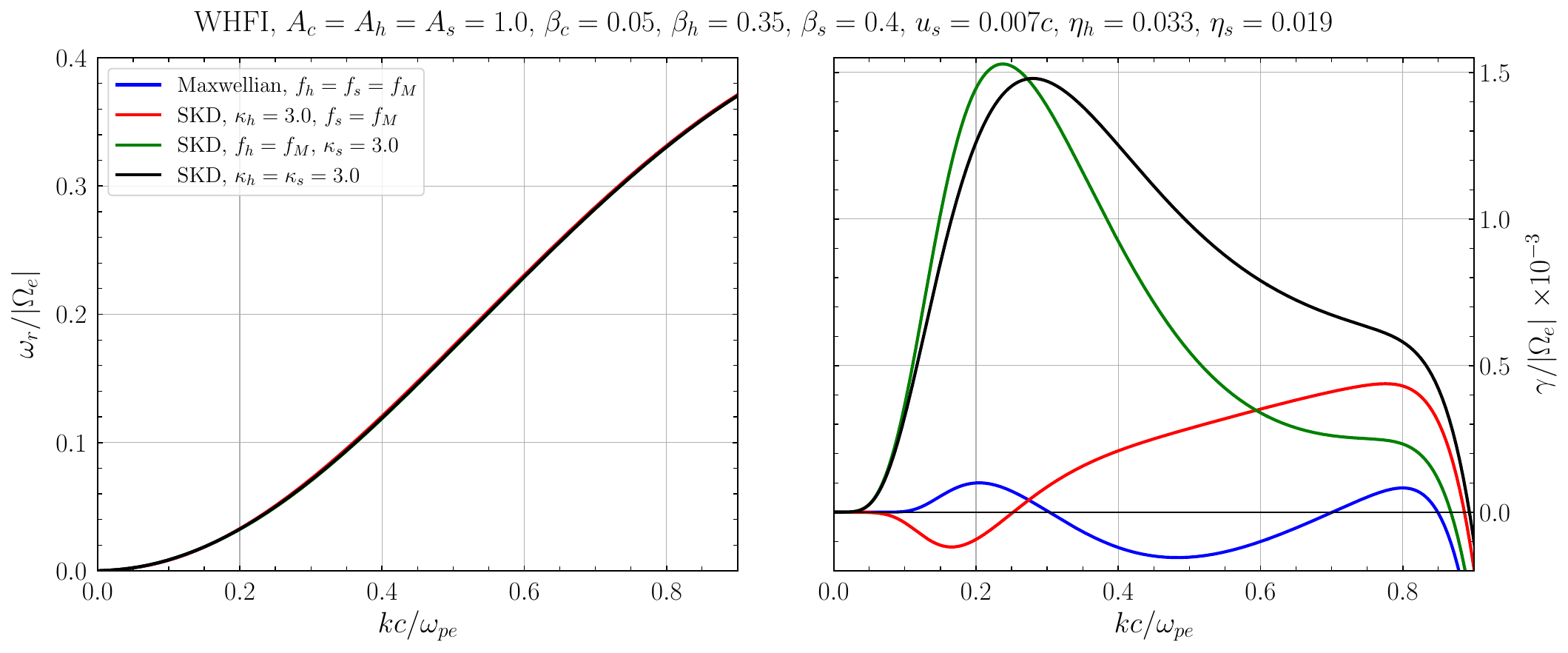}
\caption{WHFI solutions derived with DIS-K for a Maxwellian core and different halo-strahl combinations (legend): Maxwellian halo and strahl (blue), SKD halo and Maxwellian strahl (red), Maxwellian halo and SKD strahl (green), and SKD halo and strahl (black). 
The growth rates, $\gamma/|\Omega_e|$ (right), show two peaks of instabilities with the same RH polarization indicated by the sign of $\omega_r/|\Omega_e| > 0$ (left). The parameters used are indicated in the title.}\label{f8}
\end{figure*}

We next adjusted the plasma parameters, as shown in the right panel of Fig.~\ref{f1}, to excite two distinct WHFIs. 
The resulting dispersion curves, $\omega_r/|\Omega_e| > 0$ (left) and $\gamma/|\Omega_e|$ (right), are displayed in Fig.~\ref{f8}. 
In the reference case, in which both the halo and strahl are Maxwellian (blue), two RH-polarized peaks emerge: the first at lower wave numbers with  $\gamma_{\max}/ |\Omega_e| \approx 1.00\times 10^{-4}$ at $kc/\omega_{\mathrm{pe}} \approx 0.204$, driven by the strahl–halo drift; and the second, slightly weaker peak at higher wave numbers associated with the strahl–core drift, with $\gamma_{\max}/|\Omega_e| \approx 8.27\times 10^{-5}$ at $kc/\omega_{\mathrm{pe}} \approx 0.799$. 
Waves are damped across the intermediate wave-number range. 

When suprathermal tails are added only to the halo ($\kappa_h = 3.0$), the first mode becomes fully damped analogously to the WHFI behavior seen in Fig.~\ref{f6}, indicating that the strahl–halo WHFI is particularly sensitive to the temperature contrast between these two populations. 
The second peak, in contrast, is noticeably amplified and broadened, extending over a wider wave-number interval and reaching $\gamma_{\max}/|\Omega_e| \approx 4.38\times 10^{-4}$ at $kc/\omega_{\mathrm{pe}} \approx 0.775$. 
When only the strahl has SKD ($\kappa_s = 3.0$), the first growth-rate peak is strongly enhanced, becoming both higher and significantly broader and reaching $\gamma_{\max}/|\Omega_e| \approx 1.53\times 10^{-3}$; this is an increase of more than an order of magnitude at a slightly higher wave number of $kc/\omega_{\mathrm{pe}} \approx 0.237$. 
The second peak is also amplified, though less strongly (only a $2.75$ increase of the maximum growth rate) than in the $\kappa_h = 3.0$ case. 
The two branches begin to accumulate, and hence overlap, extending the overall unstable wave-number interval and making the second peak less distinct. 
In the last case, when both populations were modeled with an SKD, a cumulative effect analogous to the WHFI behavior in Fig.~\ref{f6} becomes obvious. 
The first peak is significantly enhanced due to the suprathermal tails of the strahl, but the overall enhancement is weaker than in the previous case when only the strahl had an SKD. 
The temperature ratio between the strahl and halo is now lower, and this results in a slightly lower maximum growth rate of $\gamma_{\max}/|\Omega_e| \approx 1.48\times 10^{-3}$ at $kc/\omega_{\mathrm{pe}} \approx 0.278$. 
The second peak, however, is strengthened (compared to previous cases with alternative SKDs) and merges more visibly with the first peak, making it less pronounced.

\begin{figure*}[ht!] \centering
\includegraphics[width=0.9\textwidth]{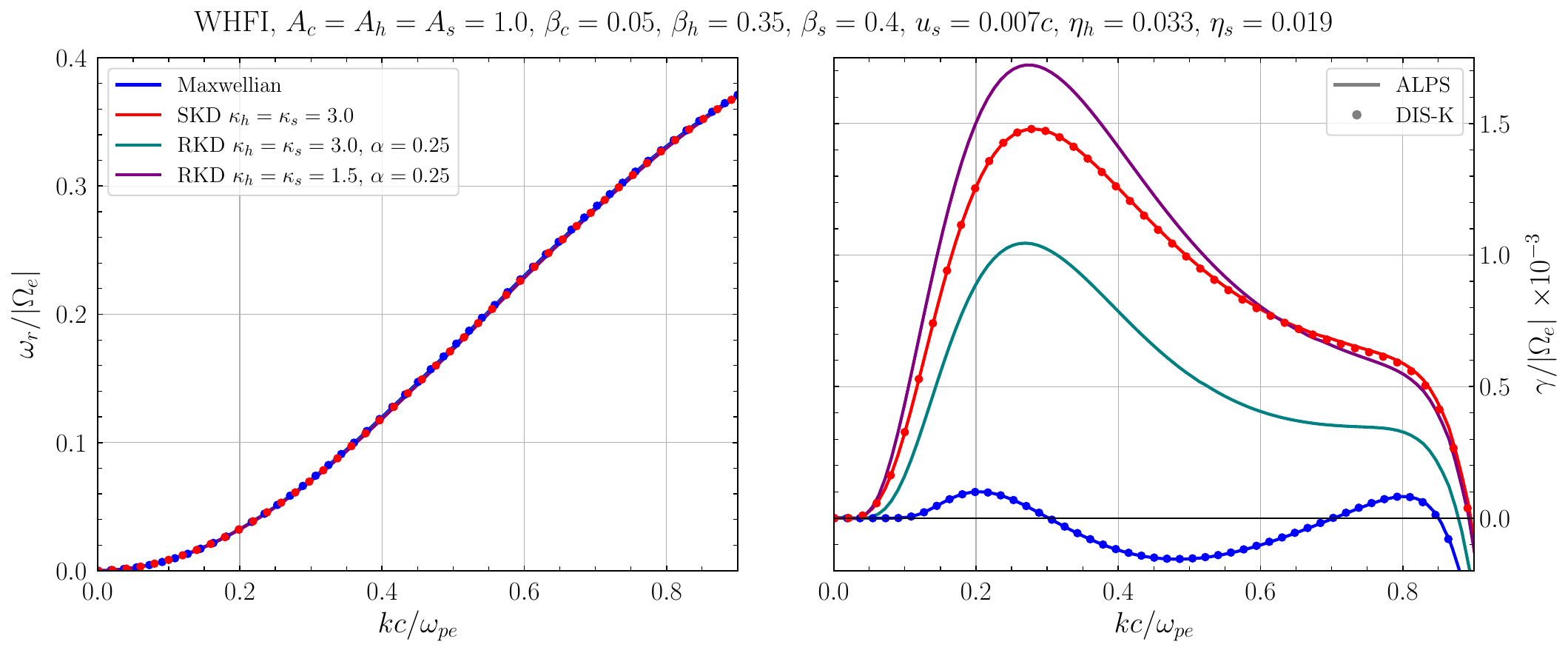}
\caption{Comparison of WHFI solutions; frequencies, $\omega_r/ |\Omega_e|$ (left); and growth rates, $\gamma/|\Omega_e|$ (right) as functions of wave number, $kc/\omega_{pe}$, obtained with DIS-K (dots) and ALPS (solid lines) for the reference case in Fig.~\ref{f8}, including the halo and strahl with Maxwellian (blue) and SKDs (red), and two additional configurations with the RKD halo and strahl populations: ($\alpha_h = \alpha_s=0.25$), $\kappa_h=\kappa_s=3.0$ (green), and $\kappa_h=\kappa_s=1.5$ (purple). 
The parameters used are indicated in the title. The agreement between DIS-K and ALPS solutions is very good.}\label{f9}
\end{figure*}

We also provide an RKD setup for these regimes with two WHFI peaks. 
Figure~\ref{f9} compares the DIS-K results (dots) with the ALPS solutions (solid lines) derived for the same reference setup in Fig.~\ref{f8}, namely, when both the halo and strahl are Maxwellian (blue) and when both have SKDs (red, $k_h=k_s=3$).
In addition, these are compared with the unstable WHFIs obtained for two configurations using RKDs for both the halo and strahl populations with $\alpha_h=\alpha_s=0.25$,   $\kappa_h=\kappa_s=3.0$ (green), and $\kappa_h=\kappa_s=1.5$ (purple).
The two solvers again show excellent agreement. 
Using RKDs with the same $\kappa$ as the SKD, but with a (finite) cutoff of the suprathermal tails ($\alpha =0.25$) reduces the suprathermal effects, making the distribution more Maxwellian, and decreases the overall growth rate notably. 
The RKD solutions derived for $\kappa_h = \kappa_s = 1.5$ are quite special as these values are not usually seen for SKDs.
In this case, more suprathermal particles are added to the system, thus increasing the first peak compared to the case with the SKD, and showing that these models could underestimate growth rates.

\section{Conclusions and outlook}\label{sec:conclusion}
%%%%%%%%%%%%%%%

{In this study, we upgraded the analysis of heat-flux instabilities, namely FHFI and WHFI, to more complex three-component electron models, typical of space plasmas in the solar wind and planetary environments.} 
These combine a Maxwellian core and two Kappa-distributed components, the so-called halo and the drifting strahl. {The new approach marks an advancement over previous models that typically considered only two electron components, such as core-beam/strahl or core-halo systems.
Our results demonstrate that accounting for all three electron components reveals new instability regimes of increased complexity that simplified two-component models fail to capture.}

We first identified two peaks of LH-polarized FHFIs. 
While the peak at lower wave numbers is driven by the relative core-strahl drift and remains little affected by the suprathermal (SKD) tails, the peak at higher wave numbers is caused by the halo-strahl drift and is highly influenced by the suprathermal tails of VDs. 
Moreover, both these two FHFI modes are highly sensitive to the strahl drift. 
Reducing the drift speed progressively weakens and shifts both peaks until they merge into a single broad maximum accompanied by a clear narrowing of the unstable wave-number interval. 
Increasing the drift instead causes the two peaks to separate: the low-wave-number peak sharpens and grows, while the high-wave-number peak shifts to larger wave numbers and decreases in amplitude until it eventually vanishes.
A transitory regime can also be identified \citep{Shaaban_mnras2018}, with two peaks of different natures: a low-wave-number LH-polarized FHFI peak, followed by a RH-polarized WHFI peak at higher wave numbers. Introducing suprathermal (SKD) tails leaves the FHFI essentially unchanged, confirming its core-strahl-driven nature, but completely suppresses the WHFI. 
The later is enhanced by the halo–strahl temperature contrast. 
Adding SKD tails to the strahl instead reverses the trend; the FHFI weakens, while the WHFI becomes stronger and broader, expanding the unstable range.
When both halo and strahl follow SKDs, the FHFI again remains largely unaffected, whereas the WHFI is still enhanced, but less strongly than with an SKD strahl alone, reflecting the reduced halo-strahl temperature ratio.

In the setup with two WHFI peaks, the Maxwellian reference case exhibits two RH-polarized modes: the first low-wave-number peak is driven by the halo-strahl drift, while the second higher wave-number peak is triggered by the core-strahl drift. 
When only the halo has an SKD and the strahl remains Maxwellian, the first instability is entirely suppressed, demonstrating its sensitivity to the halo–strahl temperature contrast, while the higher wave-number peak is enhanced and broadened.
Adding suprathermal tails solely to the strahl has the opposite effect: the WHFI peak at a low wave number becomes stronger and significantly wider, and the high-wave-number peak is also enhanced, although to a lesser extent. 
When both halo and strahl follow SKDs, a cumulative behavior of both cases is noticed: the first peak is amplified, but slightly less than for an SKD strahl alone, owing to the reduced temperature contrast, while the second peak becomes higher than in the previous cases.

For the cases analyzed here, it can be concluded that when the two peaks are of the same nature, either when they are both FHFIs or both WHFIs, the new instability triggered by the halo-strahl drift can significantly compete with the one driven by the core-strahl drift invoked in previous studies (with two-component electron systems).
Note the case of WHFI, where the maximum growth rate of the new instability can be several times higher, especially in the presence of suprathermal tails.
We also show that for systems with RKDs (for which analytical derivation of dispersion relations remains a challenge), ALPS can be successfully applied.
It thus becomes possible to consistently describe all the observed populations, including those with harder suprathermal tails (associated with $\kappa \leqslant 3/2$) incompatible with the SKDs used so far in observational analysis.

{Currently, direct observational identification of such coupled branches of enhanced fluctuations could be difficult, especially in the case of FHFI, as mentioned in the introduction.
However, we can already suggest that the analysis of narrowband emissions could be refined to correlate with predictions such as ours. 
For example, we can expect that events with simultaneous (multiple) emissions of different frequencies, but still close to each other (see Fig.~6 in \cite{Jagarlamudi-etal-2020} or Fig.~9 in \cite{Jagarlamudi-etal-2021}) could find their explanation in multiple micro-sources as in our upgraded model.
Our results therefore provide concrete theoretical predictions that can be tested against such observations in the future.
Even if individual branches are not directly resolved observationally, the unstable three-component regimes revealed here may influence the global regulation of electron heat flux.
The additional interplay of instabilities emerges from physically consistent solutions of the kinetic dispersion relation under realistic multi-component conditions in space plasmas.} However, we introduced several limitations to make the new complex analysis more transparent.
First, the assumption of isotropic temperature excludes anisotropy-driven instabilities. 
Second, by assuming zero halo-core drift, we omitted a class of heat-flux instabilities that could be generated in space plasmas when this drift is noticeable.
Third, the restriction to parallel propagation neglects oblique whistler modes that may dominate heat-flux regulation. 
Finally, the study is purely linear and does not address nonlinear wave–particle interactions.
Despite these limitations, the present results provide a systematic exploration of three-component effects within kinetic linear theory.
Moreover, our results should motivate future studies to reconsider these instabilities and their interaction in quasilinear theories and numerical simulations {and may offer valuable predictions to be tested by recent or future spacecraft observations}.
With the aim of reaching realistic conclusions regarding the role of these instabilities in regulating the heat flux mainly transported by the electron halo and strahl populations in the solar wind, {future work
should extend the linear analysis to oblique propagation --including the effects of anisotropic temperature-- and to quasilinear and numerical models to simulate the time evolution of these instabilities.}

%%%%%%%%%%%%%%%%%%%%%%%%%%%%%%%%%%%%%%%%%%%%%%%%%%%%%%%%%%%%%%

% 
\newpage

%%%%%%%%%%%%%%%%%%%%%%%%%%%%%%%%%%%%%%%%%%%%%%%%%%%%%%%%%%%%%%

%%%%%%%%%%%%%%%%%%%%%%%%%%%%%%%%%%%%%%%%%%%%%%%%%%%%%%%%%%%%%%
\begin{acknowledgements}
      The authors acknowledge support from the Ruhr-University Bochum, the Katholieke Universiteit Leuven, and the use of the ALPS code. This project was funded by the Deutsche Forschungsgesellschaft (DFG), project FI $706/31$-$1$ and the Belgian FWO-Vlaanderen G$002523$N, and SIDC Data Exploitation (ESA Prodex), No. $4000145223$. R.A.L. acknowledges the support of ANID Chile through FONDECyT grant No. $1251712$. The authors gratefully acknowledge Daniel Verscharen and Kris Klein for their continuous support with ALPS, particularly for valuable discussions, assistance with technical questions, and ongoing updates to their solver.
\end{acknowledgements}

\bibliographystyle{aa}
\bibliography{bib}

\appendix

\onecolumn
\section{Dispersion and stability resolved with ALPS} \label{appendixA}
\FloatBarrier %\usepackage{placeins}
%\twocolumn
%\FloatBarrier %\usepackage{placeins}
ALPS uses the general expression of the plasma's susceptibilities \citep{Verscharen_2018}
\begin{equation}
\boldsymbol{\chi}_j=\frac{\omega_{\mathrm{p}, j}^2}{\omega \Omega_{j}} \int_0^{\infty} 2 \pi p_{\perp} \mathrm{d} p_{\perp} \int_{-\infty}^{+\infty} \mathrm{d} p_{\|}\left[\hat{\boldsymbol{e}}_{\|} \hat{\boldsymbol{e}}_{\|} \frac{\Omega_{j}}{\omega}\left(\frac{1}{p_{\|}} \frac{\partial f_{0 j}}{\partial p_{\|}}-\frac{1}{p_{\perp}} \frac{\partial f_{0 j}}{\partial p_{\perp}}\right) p_{\|}^2\right. 
\left.+\sum_{n=-\infty}^{+\infty} \frac{\Omega_{j} p_{\perp} U}{\omega-k_{\|} v_{\|}-n \Omega_{j}} \boldsymbol{T}_n\right],
\end{equation}
 
 with
 \begin{equation}
U \equiv \frac{\partial f_{0 j}}{\partial p_{\perp}}+\frac{k_{\|}}{\omega}\left(v_{\perp} \frac{\partial f_{0 j}}{\partial p_{\|}}-v_{\|} \frac{\partial f_{0 j}}{\partial p_{\perp}}\right),
\end{equation}
 
and the tensor  
\begin{equation}
\renewcommand{\arraystretch}{2.7}
\boldsymbol{T}_n \equiv
\begin{pmatrix}
\dfrac{n^{2} J_n^{2}}{z^{2}}
& \dfrac{i n J_n J_n'}{z}
& \dfrac{n J_n^{2} p_{\|}}{z p_{\perp}} \\
-\dfrac{i n J_n J_n'}{z}
& \left(J_n'\right)^{2}
& -\dfrac{i J_n J_n' p_{\|}}{p_{\perp}} \\
\dfrac{n J_n^{2} p_{\|}}{z p_{\perp}}
& \dfrac{i J_n J_n' p_{\|}}{p_{\perp}}
& \dfrac{J_n^{2} p_{\|}^{2}}{p_{\perp}^{2}}
\end{pmatrix},
\end{equation}
where $z \equiv k_{\perp} v_{\perp} / \Omega_{j}$, and $J_n \equiv J_n(z)$ denotes the $n$ th-order Bessel function. Using the plasma's dielectric tensor as 
\begin{equation}
    \boldsymbol{\epsilon} = \mathbbm{1} + \sum_j \boldsymbol{\chi}_j,
\end{equation}
where $\mathbbm{1}$ is the unity tensor, the wave equation can be written as 
\begin{equation} 
    (\boldsymbol{k}c/\omega) \times [(\boldsymbol{k}c/\omega) \times \boldsymbol{E}] + \boldsymbol{\varepsilon} \cdot \boldsymbol{E} \equiv \mathcal{D} \cdot \boldsymbol{E} = 0,
\end{equation}
with the electric field $\boldsymbol{E}$. Nontrivial solutions ($\boldsymbol{E} \ne 0$) are then obtained by solving the dispersion relations (with small $z \to 0$ for (quasi-)parallel propagation)
\begin{equation}
{\rm det}\,\mathcal{D} = 0.
\end{equation}

% %%%%%%%%%%%%%%%%%%%%%%%%%%%%%%%%%%%%%%%%%%%%%%%%%%%%%%%%%%%%%%

% %%%%%%%%%%%%%%%%%%%%%%%%%%%%%%%%%%%%%%%%%%%%%%%%%%%%%%%%%%%%%%

\end{document}